\documentclass[prb,twocolumn,superscriptaddress,amsmath,amssymb,aps,longbibliography,nofootinbib]{revtex4-2}
\usepackage[colorlinks=true,urlcolor=blue,citecolor=blue,linkcolor=blue]{hyperref}
\usepackage{amsmath,amssymb,amsthm}
\usepackage{caption}
\usepackage{float}
\usepackage{graphicx,subfigure}
\usepackage{hyperref}
\usepackage{tikz}
\usepackage{url}
\usepackage{adjustbox}
\usepackage[capitalise]{cleveref}
\usepackage{newtxmath}
\usepackage{bm}
\usepackage[utf8]{inputenc}

\hypersetup{hypertex=true,colorlinks=true,linkcolor=blue,anchorcolor=blue,citecolor=blue} 

\newcommand{\wang}{\adjincludegraphics[valign=c,width=0.016\textwidth]{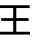}}

\begin{document}
\title{2D excitation information by MPS method on infinite helixes}

\author{Xing-Yu Zhang}
\affiliation{Institute of Physics, Chinese Academy of Sciences, Beijing 100190, China}
\affiliation{University of Chinese Academy of Sciences, Beijing 100049, China}

\author{Runze Chi}
\affiliation{Institute of Physics, Chinese Academy of Sciences, Beijing 100190, China}
\affiliation{University of Chinese Academy of Sciences, Beijing 100049, China}

\author{Yang Liu}
\affiliation{Institute of Physics, Chinese Academy of Sciences, Beijing 100190, China}
\affiliation{University of Chinese Academy of Sciences, Beijing 100049, China}

\author{Lei Wang}
\affiliation{Institute of Physics, Chinese Academy of Sciences, Beijing 100190, China}
\affiliation{Songshan Lake Materials Laboratory, Dongguan, Guangdong 523808, China}
\begin{abstract}
    Understanding the excitation spectrum in two-dimensional quantum many-body systems has long been a formidable challenge. In this study, we propose an innovative approach by introducing an excitation ansatz based on an infinite matrix product state (MPS) with a helix structure. The use of infinite uniform MPS states allows us to accurately extract key properties such as energy, degeneracy, spectrum weight, and scaling behavior of low-energy excited states simultaneously. To validate the effectiveness of our method, we apply it to the critical point of the transverse-field Ising model. The scaling exponent of the energy gap extracted aligns closely with conformal bootstrap results. Subsequently, we extend our method to the $J_1$-$J_2$ Heisenberg model on a square lattice. Our findings reveal that the degeneracy of lowest-energy excitations serves as a reliable metric for distinguishing different phases. The phase boundaries identified by our method are consistent with previous findings. The present method provides a promising avenue for studying the excitation spectrum of two-dimensional quantum many-body systems.
\end{abstract}
\maketitle

\section{Introduction}
\label{sec: introduction}
Low-energy excitations play a pivotal role in characterizing the distinctive properties of quantum many-body systems. However, the computation of excited states in quantum many-body systems remains a formidable challenge. Numerous methods have been developed to address this complex problem. Among them, the tensor network method stands out as a powerful tool for studying excitations~\cite{PhysRevLett.75.3537, PhysRevB.85.100408, PhysRevLett.121.107202, PhysRevB.101.195109, 10.21468/SciPostPhys.12.1.006, PhysRevB.104.115142, PhysRevLett.129.227201}. Nevertheless, when applying finite density matrix renormalization group (DMRG)~\cite{PhysRevLett.69.2863, PhysRevB.48.10345} and projected entangled pair states (PEPS)~\cite{verstraete2004renormalization} techniques, the absence of translation symmetry poses challenges in assigning momentum as a reliable quantum number for labeling excitations. To overcome this limitation, the single-mode approximation is employed in methods like iMPS and iPEPS calculations for infinite 1D~\cite{PhysRevLett.75.3537, PhysRevB.85.100408} and 2D~\cite{PhysRevB.92.201111, PhysRevB.101.195109, 10.21468/SciPostPhys.12.1.006, PhysRevLett.129.227201} systems, respectively. Although the graphical summation of the excitation state with the momentum~\cite{PhysRevB.85.100408, PhysRevB.101.195109, 10.21468/SciPostPhys.12.1.006} can be simplified by the derivative with generating function in finite 1D system~\cite{PhysRevB.103.205155} and infinite 2D system with discrete momentum using large unit cell~\cite{ponsioen2023improved, tu2023generating}, it is still a cumbersome matter for arbitrary geometry with continuous momentum.

While simulating an infinite 2D model directly using iPEPS may appear straightforward, obtaining the ground state even with state-of-the-art automatic differentiation optimization techniques remains challenging~\cite{10.21468/SciPostPhys.10.1.012}. Additionally, the single-mode approximation in 2D suffers from nonorthogonality of the excitation ansatz, resulting in numerical singularities when solving the eigenequation to obtain the excitation states~\cite{10.21468/SciPostPhys.12.1.006}. On the other hand, the 1D iMPS method does not encounter these issues due to the high precision with DMRG-type methods to compute the ground state and the inherent orthogonality of the excitation state facilitated by the elegant canonical form. However, recovering 2D momentum when simulating 2D systems in a 1D chain is not a trivial task. In the cylinder geometry, when placing the $y$ direction in the width direction, the $y$ direction lacks translation invariance due to the presence of a large unit cell. Therefore, to recover the $k_y$ momentum, one must solve the restricted eigenequation, which includes a hyper-parameter for numerical stability~\cite{PhysRevB.104.115142}.

In this paper, we propose an approach that employs the single-mode approximation on an infinite helix geometry to obtain the low-energy excitation spectrum of 2D systems. Compared to the iPEPS ansatz, the present approach mitigates issues related to nonorthogonality and the generalized eigenvalue solving problems, enabling us to extract excitation energy with better accuracy. In comparison to the cylinder geometry~\cite{PhysRevB.104.115142}, the ground state on the helix exhibits uniformity, requiring only a unit cell with one site. Additionally, the excitation ansatz exhibits exact 2D horizontal and vertical translation invariance, making the $k_y$ momentum a reliable quantum number and the graphical summation of the excitation energy easier to implement.

We validate the present method through two model benchmarks and achieve consistent results for phase transitions using both the ground state and excitation states within a unified framework. For transverse field Ising, we obtain comparable results of energy gap with iMPS on cylinders~\cite{PhysRevB.104.115142} and iPEPS calculations. We also calculate the scaling exponent of the energy gap at the critical point. Using critical level crossings analysis~\cite{OKAMOTO1992433, PhysRevLett.121.107202, PhysRevX.11.031034} for the $J_1$-$J_2$ model, we identify the critical point of the QSL to VBS phase transition around $J_2 = 0.52$. This critical point is found to be insensitive to variations in the system size and bond dimension. Furthermore, we directly observe the presence of four-fold degenerate lowest-energy excitations without restricted symmetry during this transition suggesting the emergent symmetry at this transition point.

The paper is organized as follows. In~\cref{sec: Method}, we provide an overview of the methods employed, including the ground state iMPS simulation in the thermodynamic limit and the single-mode approximation on the helix.~\cref{sec: Results} focuses on the application of these methods. We present results obtained from the transverse field Ising and $J_1$-$J_2$ model. In~\cref{sec: Discussion}, we discuss potential issues and directions for future research.

To facilitate practical implementation, we have developed a Julia package that provides ready-to-use functionality for the single-mode approximation on the infinite helix~\cite{AD_Excitation}.

\section{Method}
\label{sec: Method}

\begin{figure}[H]
    \centering 
    \includegraphics[width=0.4\textwidth]{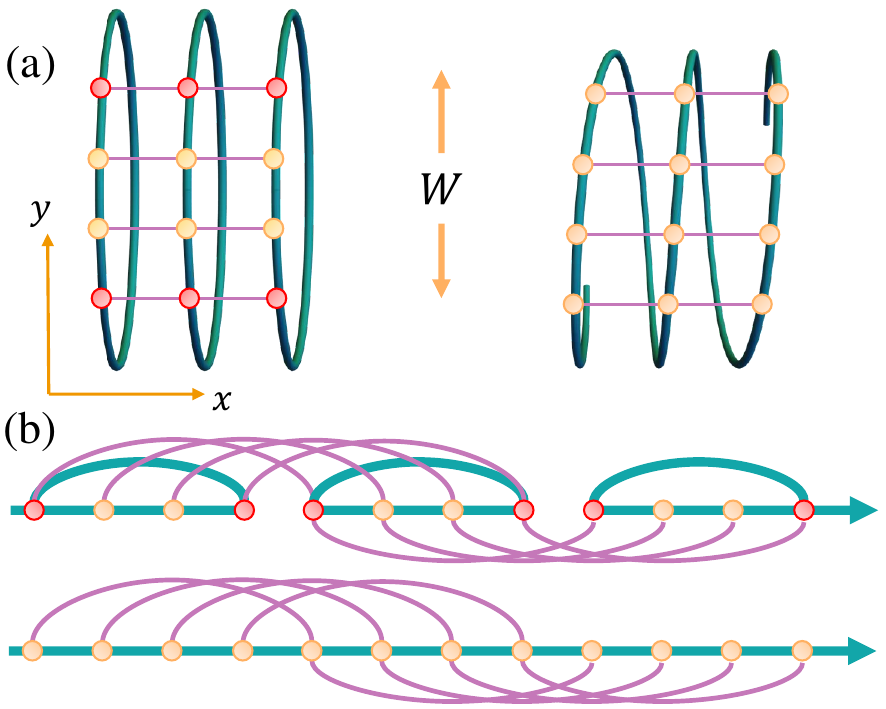}
    \caption{2D system with the nearest neighbor interaction transfer to $W=4$(Width) 1D system as an example. (a) shows the different boundary conditions when placing a 2D square lattice on the cylinder and helix respectively. Both geometries are infinite in the $x$ direction and finite in $y$ direction as labeled. (b) shows the difference in the MPO connections between the two boundary conditions. The unit cell on the cylinder is $W=4$ sites, while the unit cell on the helix is only one site.}
    \label{fig: Cylinder_and_Helix}
\end{figure}

We adopt the single-mode approximation~\cite{PhysRevB.85.100408} as the excitation ansatz. This method starts with an infinite uniform MPS serving as the ground state, which is obtained through the variational uniform matrix product state (VUMPS) method~\cite{PhysRevB.97.045145, 10.21468/SciPostPhysLectNotes.7}. We employ the infinite DMRG method~\cite{mcculloch2008infinite} to initialize the iMPS before applying VUMPS. For the present simulations, we utilize helix boundary conditions~\cite{PhysRevB.107.L081104}. The main distinction from the cylinder geometry~\cite{PhysRevB.104.115142} is the unit cell size. On the cylinder, a large unit cell is required for the matrix product operator (MPO)~\cite{SCHOLLWOCK201196}, which represents the Hamiltonian. In contrast, the helix requires only a unit cell with one site, as depicted in Fig.\ref{fig: Cylinder_and_Helix}.

After obtaining the ground state, the excitation ansatz on the helix can be expressed as:
\begin{equation}
    \begin{aligned}
    \left|\Psi_{k_x, k_y}(B)\right\rangle=&\sum_n e^{-i k_x\lfloor n / W\rfloor-i k_y(n \% W)} \times\\[-15pt]
    &\includegraphics[width=0.35\textwidth]{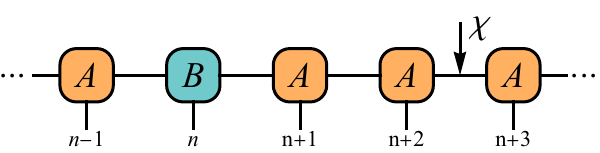} 
    \end{aligned}
    \label{equ: excitation_ansatz}
\end{equation}

Here, $\lfloor \rfloor$ denotes the floor integer operation, $\%$ represents the modulo operation. The helix has infinite length in the $x$ direction but finite width in the $y$ direction, so the $k_x$ has continuous values while $k_y=2\pi i / W\ (i=0,1,2...)$ has discrete values. It's worth noticing that the $k_x$ and $k_y$ are the reliable quantum numbers of 2D translation operator $T_x$ and $T_y$ separately~\footnote{An alternative ansatz can be written without the floor integer for the $k_x$ in~\cref{equ: excitation_ansatz}, which is translation invariance along the helixes but not in the 2D context. When $k_x=0$, the two ansatz become the same; the excitation ansatz is the eigenstates of both helix and 2D Hamiltonian. Notably, the key results presented in this paper were calculated under $k_x=0$. For general $k_x$, the difference between these two ansatz will vanish with increasing width $W$.}. The virtual bond dimension is defined as $\chi$.

To obtain the $B$ tensor, we solve the eigenequation by summing the energy contributions with an orthogonal parameterization of the excitation state~\cite{PhysRevB.85.100408, 10.21468/SciPostPhysLectNotes.7}. The graphical summation of the excitation energy on the helix is the only non-standard procedure involved in the calculation. This process consists of a series of graphical summations to calculate the excitation energy, which can be classified into three categories based on the placement of the bar $\bar{B}_m$ and the ket $B_n$ ($\bar{B}$ denotes the conjugate of $B$): (i) when $m=n$ and they are on the same site; when they are on different sites (ii) $m>n$ and (iii) $m<n$. These three terms can be represented by the corresponding tensor diagrams with canonical form $A_L$ and $A_R$ and MPO tensor $M$:

\begin{widetext}
    \begin{equation}
        \begin{aligned}
        \frac{1}{N} \sum_{n,m} \left\langle\Psi_{k_x, k_y}(\bar{B}_m)\right| H \left|\Psi_{k_x, k_y}(B_n)\right\rangle =\quad&  \adjincludegraphics[valign=c,height=0.05\textheight]{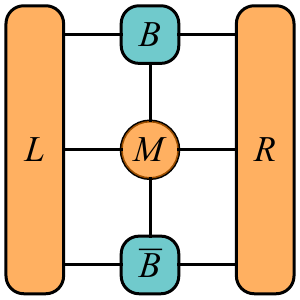} +\\
        \frac{W-1+e^{ik_x}}{W}e^{ik_y}\cdot&\adjincludegraphics[valign=c,height=0.05\textheight]{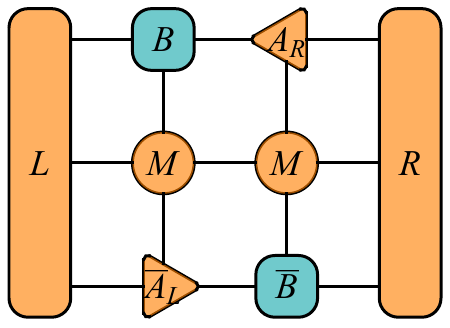} + 
        \frac{W-2+2e^{ik_x}}{W}e^{ik_y\cdot2}\cdot\adjincludegraphics[valign=c,height=0.05\textheight]{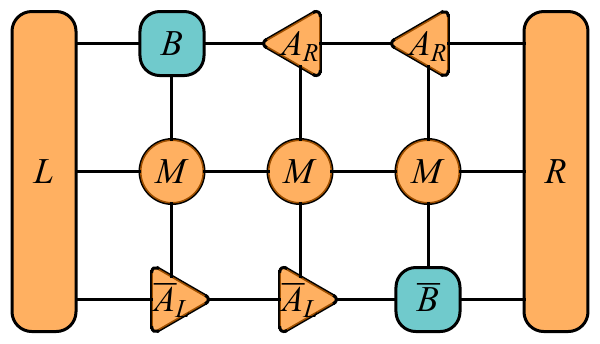} + \cdots +\\
        \frac{W-1+e^{-ik_x}}{W}e^{-ik_y}\cdot&\adjincludegraphics[valign=c,height=0.05\textheight]{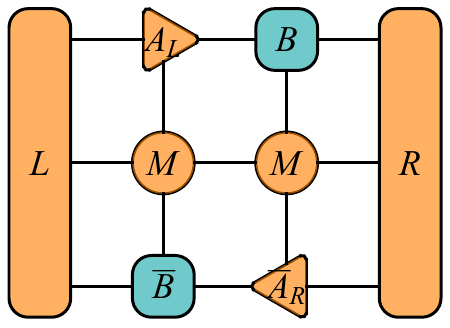} + 
        \frac{W-2+2e^{-ik_x}}{W}e^{-ik_y\cdot2}\cdot\adjincludegraphics[valign=c,height=0.05\textheight]{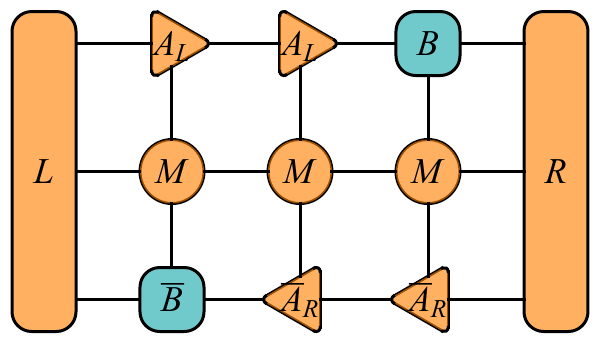} + \cdots \\
        \equiv\quad&\adjincludegraphics[valign=c,height=0.05\textheight]{./excitation_energy_on_site.pdf} + \adjincludegraphics[valign=c,height=0.05\textheight]{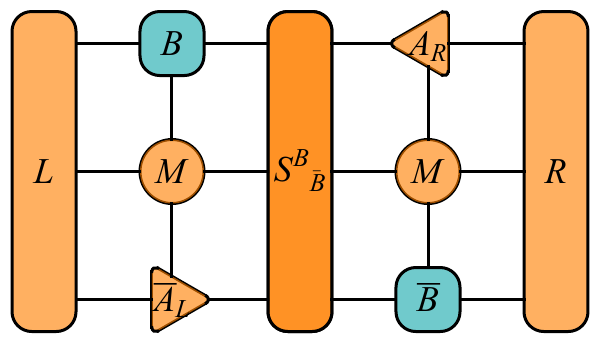} + \adjincludegraphics[valign=c,height=0.05\textheight]{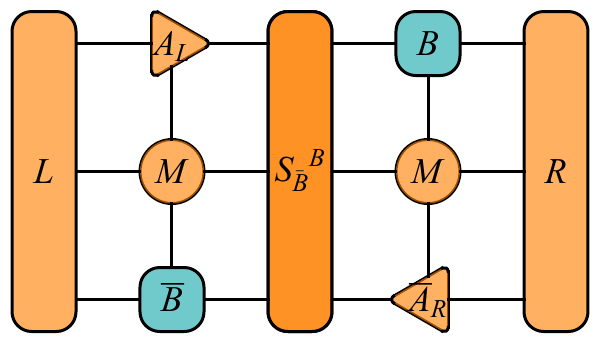}
        \end{aligned}
        \label{equ: approximation_summation}
    \end{equation}
\end{widetext}
$S^{B}_{\ \bar{B}}$ and $S^{\ B}_{\bar{B}}$ are a series summation with coefficient. In our notation, the distinctions between $S^{B}_{\ \bar{B}}$ and $S^{\ B}_{\bar{B}}$  are crucial. Specifically, $S^{B}_{\ \bar{B}}$ denotes the placement of $\bar{B}$ at the right bottom of the $B$ tensor, while $S^{\ B}_{\bar{B}}$ signifies the positioning of $\bar{B}$ at the left bottom of the $B$ tensor. These notational differences correspond to distinct summations based on their respective locations. The exact $S^{B}_{\ \bar{B}}$ can be expressed as:

\begin{equation}
    S^{B}_{\ \bar{B}} = \sum_{j=1}^W\left(\frac{W-j+je^{ik_x}}{W}e^{ik_y\cdot j}\wang^{j-1}\right)\cdot \sum_{j=0}^\infty\left(e^{ik_x}\wang^W\right)^j
    \label{equ: exact_summation}
\end{equation}
where $W$ is the width of the helix, $N$ is the total number of sites and $\wang$ is the six-order tensor transfer matrix \adjincludegraphics[valign=c,width=0.025\textwidth]{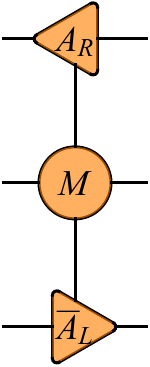} .For $m<n$, the summation $S^{\ B}_{\bar{B}}$ is just transformation of $\left(k_x, k_y\right) \rightarrow-\left(k_x, k_y\right)$ with a different transfer matrix. 

The coefficients $\frac{W-j+je^{ik_x}}{W}$ arise from two distinct types of connections between the two tensors. The first type corresponds to connections within the same column, where the coefficient is equal to $1$. In this case, there are $W-j$ possibilities contributing $\frac{W-j}{W}$. The second type corresponds to connections between different columns, where the coefficient is $e^{ik_x}$. Here, there are $j$ possibilities contributing $\frac{je^{ik_x}}{W}$. To understand this equation, let us assume $k_x=0$, in which case the summation becomes $\sum_{j=1}^W\left(e^{ik_y\cdot j}\wang^{j-1}\right)\cdot \sum_{j=0}^\infty\left(e^{ik_x}\wang^W\right)^j$. The first and second terms represent the summation of finite $y$ and infinite $x$ direction, respectively, which explains why $k_y$ and $k_x$ are summed separately. In the generic case where $k_x \neq 0$, the term $e^{ik_x}$ in the summation of $e^{ik_y\cdot j}\wang^{j-1}$ arises from the long-range connection of the nearest neighbor column on the helix, and represents the finite size effect in the present simulation due to the finite width of the helix. To mitigate the finite size effect, we propose a strategy in~\cref{sec: Mitigate the finite size effect}.

\section{Results}
\label{sec: Results}

\subsection{Transverse field Ising model}
\label{subsec: Transverse field Ising model}
\begin{figure}[H] 
    \centering
    \includegraphics[width=0.45\textwidth]{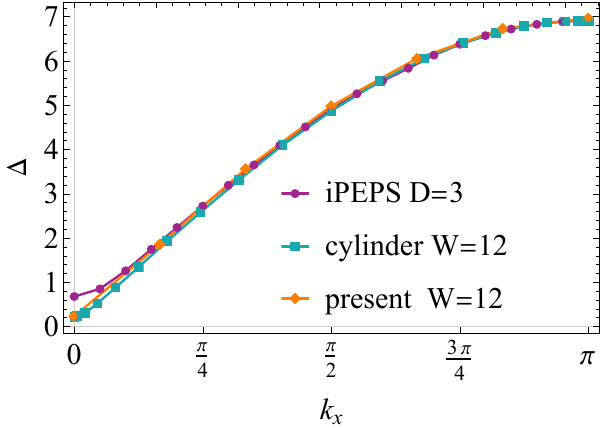}
    \caption{The gap $\Delta$ varies $k_x$ given $k_y=0$ in the transverse Ising model with $\lambda=3.04438$ ($W=12\ \chi=256$). The present results exhibit consistency with the method on the cylinder geometry~\cite{PhysRevB.104.115142} and iPEPS calculations. At the $\Gamma(0,0)$ point, the gap derived from the MPS method is comparatively smaller than that obtained from the iPEPS method.}
    \label{fig: TFIsing_gap-kx_approx}
\end{figure}

To validate the present method, we conduct a benchmark study on the 2D transverse-field Ising model, described by the Hamiltonian:
\begin{equation}
    H=-\sum_{\langle i, j\rangle} \sigma_i^z \sigma_j^z-\lambda \sum_i \sigma_i^x
\end{equation}
Here, $\sigma$ represents the spin-1/2 Pauli operator and $\langle i, j\rangle$ denotes the nearest neighbor sites. We specifically choose the quantum critical point $\lambda=3.04438$~\cite{PhysRevE.66.066110} to calculate the excitation spectrum, as depicted in \cref{fig: TFIsing_gap-kx_approx}. 
    
\begin{figure}[H] 
    \centering
    \includegraphics[width=0.45\textwidth]{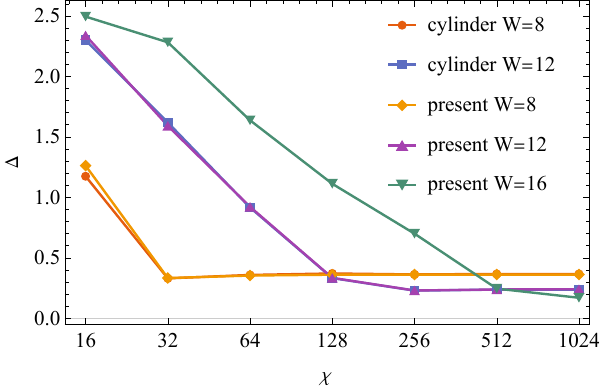}
    \caption{The energy gap $\Delta$ at the $\Gamma(0,0)$ point in the Ising model with $\lambda=3.04438$ for different system widths $W$ and MPS virtual bond dimensions $\chi$. The present results exhibit consistency with the method on the cylinder geometry~\cite{PhysRevB.104.115142}. As the virtual bond dimension $\chi$ increases, the gap converges, while the gap decreases with increasing system width $W$.} 
    \label{fig: TFIsing_gap_kx0_ky0}
\end{figure}

At the critical point, the $\Gamma(0,0)$ point in the Brillouin zone is expected to be gapless. However, in the present finite width calculations, a small gap is observed when the width $W$ is not sufficiently large. As shown in Fig.~\ref{fig: TFIsing_gap_kx0_ky0}, this gap gradually decreases as the width $W$ increases.

\begin{figure}[H] 
    \centering
    \subfigure{
        \includegraphics[width=0.465\textwidth]{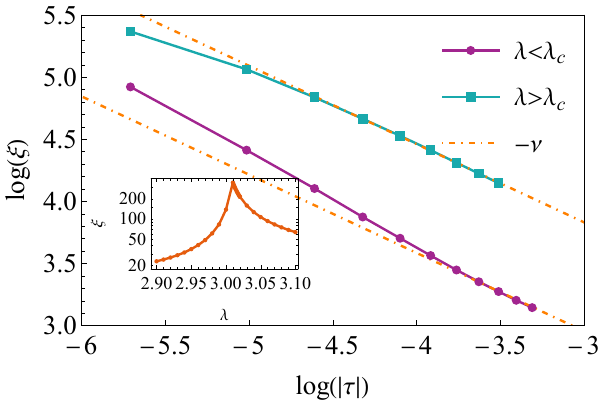}
        } 
    \subfigure{
        \includegraphics[width=0.45\textwidth]{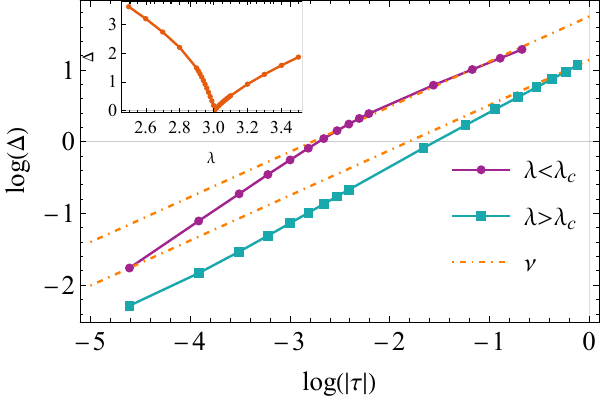}
        }
    \caption{Top: the correlation length $\xi$ as a function of $|\tau|$ for the Ising model ($W=12 \ \chi=1024$), displayed on a log-log scale with the theoretically anticipated slope of $-\nu$ \cite{Kos:2016ysd} by orange dashdot line, which indicates $\xi \propto \tau^{-\nu}$. Bottom: the gap $\Delta$ as a function of $|\tau|$ with the slope of $\nu$. The insets depict $\xi$ and $\Delta$ along $\lambda$.}
    \label{fig: TFIsing_exponent}
\end{figure}

To validate the accuracy of the present method, we calculate the critical exponent $\nu$ through the scaling behavior of correlation length $\xi$ and the gap $\Delta$ at the $\Gamma(0,0)$ point. The critical exponent $\nu$ can be extracted through the $\xi \propto \tau^{-\nu}$, where $\tau=(\lambda-\lambda_c)/\lambda_c$, and also $\Delta \propto \tau^{\nu z}$ with $z=1$ as the dynamics critical exponent in Lorentz invariant system.  We determine $\lambda_c=3.01$ in the $W=12$ case as the minimum of the gap. The correlation length, $\xi=-1 / \log(|\lambda_1 / \lambda_0|)$, is determined using the groundstate transfer matrix generated through the iMPS approach. Here, $\lambda_0$ and $\lambda_1$ represent the largest and second-largest eigenvalues of this transfer matrix. As shown in~\cref{fig: TFIsing_exponent}, the obtained critical exponent is consistent between these two schemes and in good agreement with the value reported in the most recent conformal bootstrap study, which is $\nu=0.629971(4)$~\cite{Kos:2016ysd}.

\subsection{$J_1$-$J_2$ Heisenberg model}
\label{subsec: $J_1$-$J_2$ Heisenberg model}

The two-dimensional spin-1/2 antiferromagnetic $J_1$-$J_2$ Heisenberg model on the square lattice has been the subject of extensive research for several decades due to its simplicity and yet rich physical implications~\cite{PhysRevB.38.9335, PhysRevLett.63.2148, PhysRevLett.66.1773, RevModPhys.78.17}. The Hamiltonian that characterizes this model is given by:
\begin{equation}
    H = J_1 \sum_{\langle i, j\rangle} \hat{\mathbf{S}}_i \cdot \hat{\mathbf{S}}_j + J_2 \sum_{\langle\langle i, j\rangle\rangle} \hat{\mathbf{S}}_i \cdot \hat{\mathbf{S}}_j \quad \left(J_1, J_2 > 0\right)
\end{equation}
with $J_1$, $J_2$ as the strength of the nearest neighbor and the next-nearest neighbor interaction separately. 
For the sake of convenience, we adopt the convention of setting $J_1=1$ throughout this paper. The introduction of $J_2$ has led to a rich phase diagram. In the limit of small $J_2$, the ground state exhibits N\'eel order. Conversely, for large values of $J_2$, the phase displays antiferromagnetic stripe order. In the intermediate range of $J_2$, prior research has suggested the existence of a potential quantum spin liquid (QSL) phase~\cite{PhysRevB.38.9335, doi:10.1126/science.235.4793.1196, PhysRevB.39.11413, PhysRevB.41.4619, PhysRevB.46.8206, PhysRevB.51.6151, refId0, Richter2010, PhysRevB.86.024424, PhysRevB.86.045115, PhysRevB.89.235122, PhysRevB.90.041106} within a small region adjacent to the N\'eel phase. Additionally, there is a valence-bond solid (VBS) state~\cite{PhysRevB.41.4619, PhysRevB.41.9323, PhysRevB.78.214415} located between the QSL and striped phase, with proposals including a columnar valence-bond solid state~\cite{PhysRevB.44.12050, PhysRevB.43.10970, PhysRevLett.63.2148, PhysRevB.41.9323, H.J.Schulz_1992, PhysRevB.60.7278, PhysRevB.73.184420, PhysRevB.79.195119, doi:10.7566/JPSJ.84.024720} and a plaquette valence-bond solid state~\cite{PhysRevB.54.9007, PhysRevLett.84.3173, PhysRevLett.91.197202, PhysRevB.74.144422, PhysRevB.79.024409, PhysRevB.85.094407, PhysRevB.89.104415}. Recent research also indicates that the QSL may vanish in the thermodynamic limit~\cite{qian2023absence}. In the following, we apply the present method to this model with the intricate phase diagram to extract the information of the excitation spectrum.

\begin{figure}[H]
    \centering 
    \subfigure{
        \includegraphics[width=0.225\textwidth]{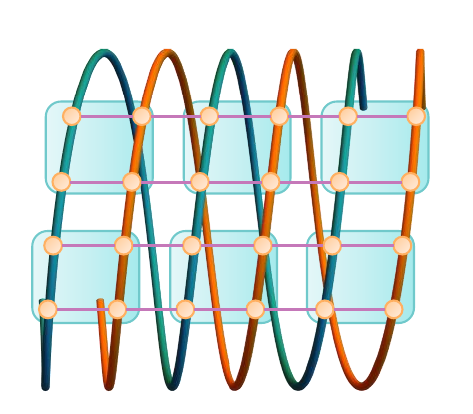}
        }
    \subfigure{
        \includegraphics[width=0.225\textwidth]{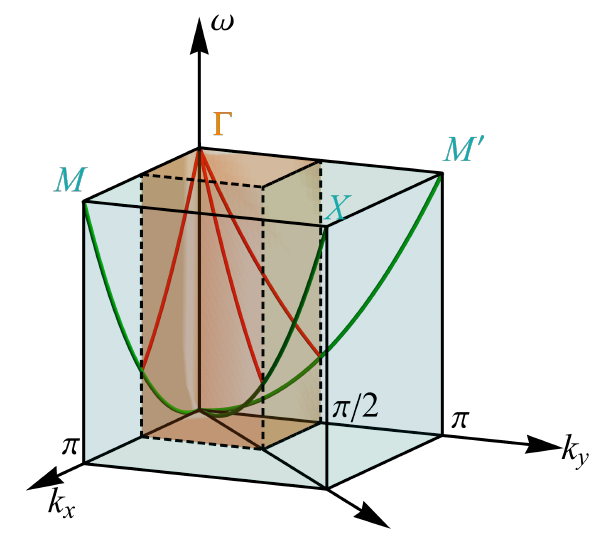}
        } 
    \caption{Left: A schematic diagram illustrating the boundary conditions achieved by merging a $2 \times 2$ unit cell on two helices. Right: Representation of Brillouin zone folding, where the original Brillouin zone is in green and the folded zone is in red.}
    \label{fig: Helix_2x2}
\end{figure}

In the present simulations, we consider a large unit cell size of $2 \times 2$. To achieve this, we merge these four individual sites into a larger unit cell. The boundary condition we take in this simulation is equivalent to two helixes as illustrated in~\cref{fig: Helix_2x2}. To ensure consistency in simulations of the $J_1$-$J_2$ model, we always use an even number for the width ($W$) of the system.

Merging of $2\times 2$ sites increases the computational cost, but it offers several advantages. Firstly, it preserves translational invariance in the helix system, simplifying the computation of excited states without the need for a large unit cell. Secondly, larger unit cells help describe more non-local excitations. Thirdly, the merging of the crucial points $\Gamma(0,0)$, $M(\pi,0)$, $M^\prime(0,\pi)$, and $X(\pi,\pi)$ into the large unit cell at $(0,0)$ enables us to obtain information about these four points by calculating just one $k$ point. Furthermore, this merged point at $(0,0)$ is precise without any approximation arising from the finite width, as described by Appendix~\cref{equ: approximation_summation}.

The lowest excitations that were originally observed at the $M$ and $X$ points now manifest as higher excitations at the present $\Gamma$ point, depicted as the red line in~\cref{fig: Helix_2x2}. However, we can still distinguish these excitations based on their spectrum weights $|\langle\Psi_{\boldsymbol{k}}(B_m^{\dagger})|O_{\boldsymbol{k}}| \Psi(A)\rangle|^2$, where $O_{\boldsymbol{k}}=\sum_{\boldsymbol{n}} O_{\boldsymbol{n}} e^{i \boldsymbol{k} \cdot \boldsymbol{n}}$ as operator of $O$-type excitation and $m$ represents the $m$-th excitation state. For instance, in the case of the Heisenberg model, we identify the gap at the momentum sector ($\pi$,0) by evaluating the transverse spin spectrum weight, where $O_{\boldsymbol{n}}$ is the spin-1/2 $S^z$ operator,  of each excitation state at ($\pi$,0) and selecting the state with the largest spectrum weight. 



\subsubsection{$J_2=0$ Heisenberg model}
\label{subsubsec: $J_2=0$ Heisenberg model}
\begin{figure}[H]
    \centering
    \includegraphics[width=0.5\textwidth]{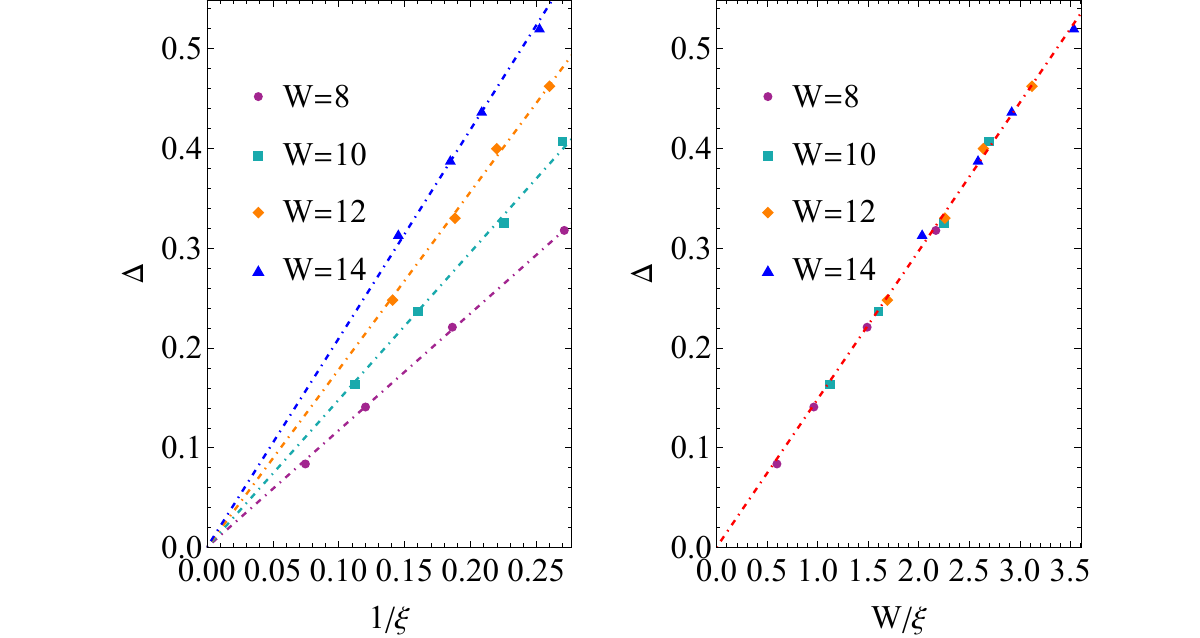}
    \caption{The energy gap $\Delta$ extrapolation at the $\Gamma(0,0)$ point for the case when $J_2=0$, using the correlation length $\xi$ ($\chi=64,128,256,512$). Left: the extrapolation is performed with the correlation length $\xi$, revealing that the gap extrapolates with different slopes for varying system widths $W$. Right:  the extrapolation is conducted with the correlation length ratio $\xi/W$, resulting in the collapse of different $W$ values onto a single line.} 
    \label{fig: J20_kx0_ky0_gap_extrapolation}
\end{figure}

\begin{figure}[H] 
    \centering
    \includegraphics[width=0.45\textwidth]{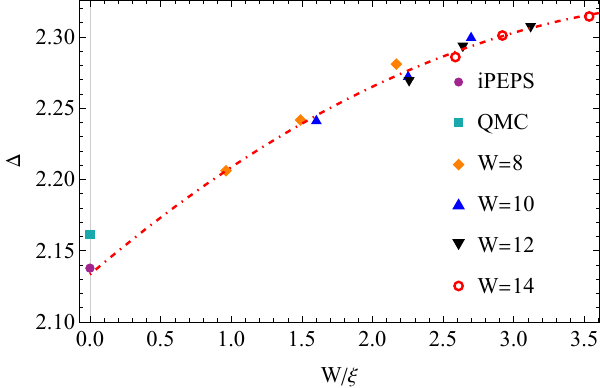}
    \caption{The energy gap $\Delta$ extrapolation at the $M(\pi,0)$ point for the case when $J_2=0$, using the correlation length ratio $\xi/W$. We employ quadratic polynomial fitting to obtain the gap values, which closely align with the results obtained from iPEPS~\cite{PhysRevB.101.195109} and QMC~\cite{PhysRevLett.86.528}.} 
    \label{fig: J20_kxpi_ky0_gap_extrapolation}
\end{figure}

We employ the $J_2=0$ limit, corresponding to the Heisenberg model, to validate the $2\times 2$ unit cell ansatz. We observe an exact zero gap at $\Gamma(0,0)$ as shown in~\cref{fig: J20_kx0_ky0_gap_extrapolation}, and a finite gap at $M(\pi,0)$ in~\cref{fig: J20_kxpi_ky0_gap_extrapolation} by extrapolating with the ratio between correlation length and width of helix, represented as $\xi/W$. This outcome contrasts with the previous finding that the spin excitation for spin-1/2 Heisenberg ladder is gapped with the even-numbered chain~\cite{Rice1993, White1994}. This discrepancy can be attributed to the helix boundary condition which breaks the short-range resonating valence bond picture. The gap at $(\pi,0)$ determined by the present method is $\Delta=2.13$, closely aligns with the results obtained from iPEPS~\cite{PhysRevB.101.195109} and QMC~\cite{PhysRevLett.86.528}.

\subsubsection{Low-Energy Excitations}
\begin{figure}[H] 
    \centering
    \includegraphics[width=0.5\textwidth]{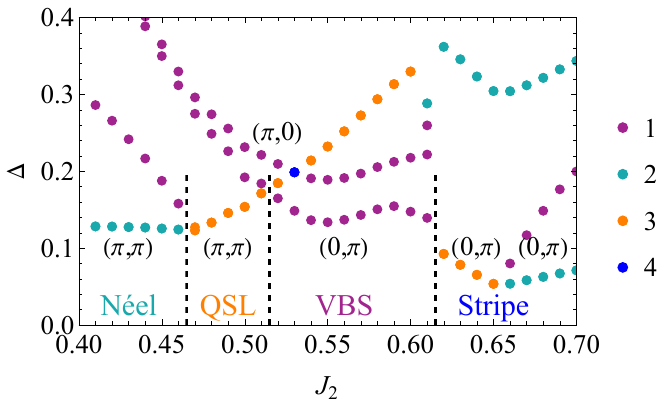}
    \caption{The energy gap $\Delta$ for excited state with varying $J_2$ ($W=10, \chi=1024$), where different colors denote the degeneracy of the excitations. The momentums of excitation states identified by spectral weight are also labeled. The degeneracy of the lowest excitation changes as $J_2$ increases. This information allows us to identify the phase transitions in the system.} 
    \label{fig: gap-J2_W5_chi1024}
\end{figure}

By analyzing the low-energy excitation spectrum, as illustrated in~\cref{fig: gap-J2_W5_chi1024}, we can identify the phase diagram of the $J_1$-$J_2$ model, encompassing four distinct phases. In the following, as increasing $J_2$, we will provide an in-depth exploration of the phase diagram, taking into consideration factors such as excitation degeneracy, spectral weight, and groundstate properties like magnetic and bond energy, as needed. 

In the Néel phase ($J_2<0.46$), the lowest and second lowest excitation states exhibit double and single degeneracy, respectively. The transverse spin spectral weight of these two degenerate excitation states is concentrated at the point $(\pi, \pi)$. These degeneracies arise from the spontaneous symmetry breaking of SU(2). 

\begin{figure}[H] 
    \centering
    \includegraphics[width=0.45\textwidth]{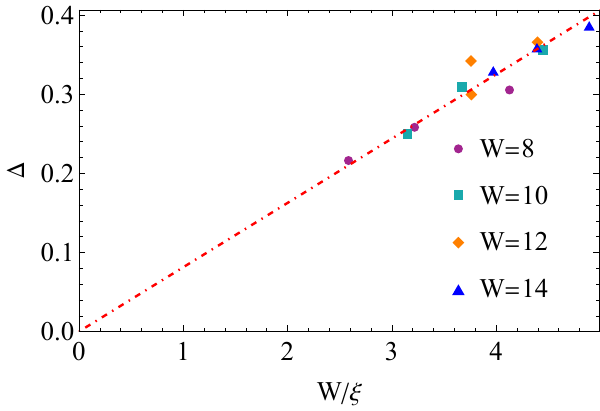}
    \caption{The energy gap $\Delta$ extrapolation at the $\Gamma(0,0)$ point for the case when $J_2=0.5$, using the  correlation length ratio $\xi/W$. } 
    \label{fig: J20.5_kx0_ky0_gap_extrapolation}
\end{figure}

In the QSL phase ($0.46<J_2<0.52$), the lowest excitation state displays a triple degeneracy at $(\pi,\pi)$, indicating the conserving of SU(2) symmetry. The transition point around 0.46 is consistent with the results obtained from the magnetic results in the thermodynamic limit as illustrated in Appendix~\cref{fig: m2-J2}. The gap extrapolate suggests that this phase is a gapless QSL, as shown in~\cref{fig: J20.5_kx0_ky0_gap_extrapolation}.

In the VBS phase ($0.52<J_2<0.61$), the lowest excitation state has a single degeneracy at $(0,\pi)$, captured by the dimer spectral weight defined as $|\langle\Psi_{\boldsymbol{k}}(B_m^{\dagger}) | D^{\boldsymbol{x}}_{\boldsymbol{k}} |\Psi(A)\rangle|^2 $, where $D_{\boldsymbol{k}}^{\boldsymbol{x}}=\sum_{\boldsymbol{n}} S_{\boldsymbol{n}}\cdot S_{\boldsymbol{n}+\boldsymbol{x}}e^{i\boldsymbol{k}\cdot \boldsymbol{n}}$ and $m$ represents the $m$-th excitation state. This indicates the staggered VBS nature of the ground state. The phase boundary between the QSL and VBS phase is characterized by the crossing of triplet and singlet around 0.52, a finding that aligns with previous research approaches~\cite{OKAMOTO1992433, PhysRevLett.121.107202, PhysRevX.11.031034}. In the 1D $J_1$-$J_2$ model, the singlet-triplet level crossing phenomenon can be attributed to the distinct scaling behavior of singlet and triplet excitations concerning the system size, a concept originally derived from conformal field theory~\cite{OKAMOTO1992433}. The present findings suggest a similar scenario in the 2D $J_1$-$J_2$ model, where we conduct simulations with finite correlation lengths. In this case, the level crossing becomes possible. Remarkably, the observed level crossing appears to exhibit insensitivity to changes in the system width as depicted in Appendix~\cref{fig: gap-J2_all}. 

\begin{figure}[H] 
    \centering
    \includegraphics[width=0.45\textwidth]{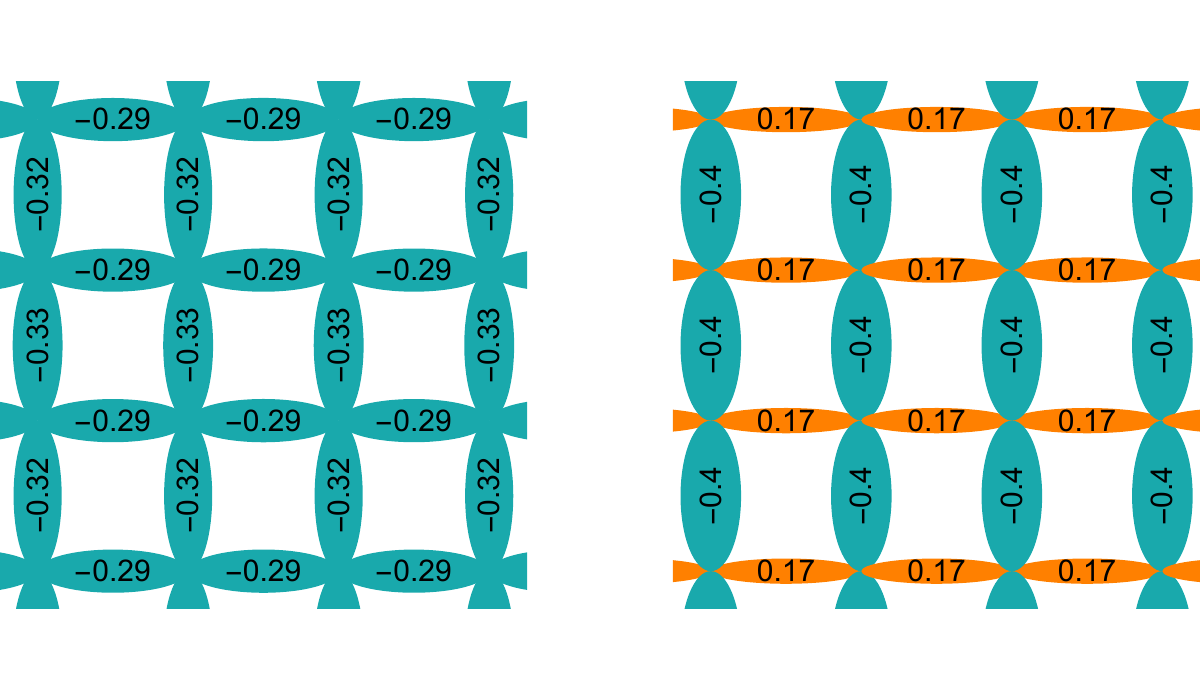}
    \caption{The bond energy of ground states ($W=10 \ \chi=1024$) at $J_2=0.5$ (Left) and $J_2=0.63$ (Right). The bond energy of ground states in the stripe phase is different from that in the QSL phase.} 
    \label{fig: bond_energy}
\end{figure}

In the stripe phase ($0.61<J_2$), the lowest excitation state is expected to exhibit a doublet degeneracy at $(0,\pi)$. The phase boundary between the VBS and stripe phase is characterized by first-order energy transition occurring around 0.61, aligning with previous research approaches~\cite{PhysRevB.86.024424, PhysRevB.88.060402, PhysRevLett.113.027201, PhysRevX.11.031034}. We find the existence of a region with three-fold lowest energy excitation depends on the choice of $W$ and bond dimension $\chi$. This three-fold degeneracy arises when $\chi$ is large and $W$ is small, and it disappears with large $W$, as depicted in Appendix~\cref{fig: gap-J2_all}. Although the lowest excitation state is triplet both in the QSL and the small region of stripe phases, the bond energies of ground states are different as shown in~\cref{fig: bond_energy}. The stripe phases break the spatial symmetry indicating that they cannot be considered QSL phases. (The slight horizontal and vertical differences of bond energy at $J_2=0.5$ are attributed to numerical precision in simulating a 2D system using a 1D chain). The emergence of the additional three-fold degenerate lowest excitation state is actually a consequence of the 1D nature of the present ansatz, which tends to preserve the SU(2) symmetry.

\section{Discussion}
\label{sec: Discussion}
In summary, utilizing the single-mode approximation on an infinite helix structure enables us to compute low-energy excitations of 2D systems. When applied to the square lattice transverse field Ising model, we obtain a rather accurate critical exponent of the energy gap near the critical point. Furthermore, we applied it to the $J_1$-$J_2$ model, we found that the degeneracy of the low-energy excitations can serve as a good probe for distinguishing different phases.

\begin{figure}[H] 
    \centering
    \includegraphics[width=0.45\textwidth]{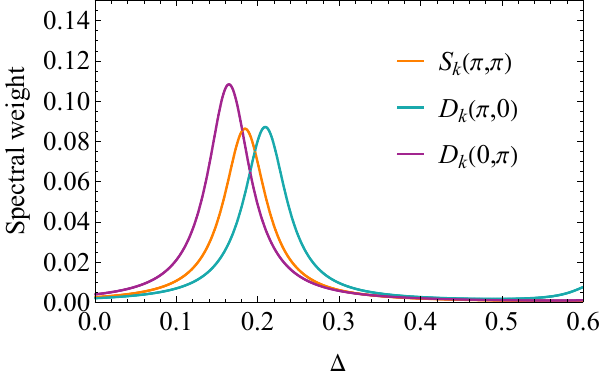}
    \caption{The figure presents the spectral weight of excitation state at $J_2=0.52$ for a system with $W=10$ and $\chi=1024$. The plot reveals the proximity of two singlet states and one triplet state of the lowest five excitation states, indicating the presence of a quintuplet state for finite $W$. All sets of data were convoluted with a Lorentzian function to account for broadening effects.} 
    \label{fig: spectral_weight_W5_chi1024}
\end{figure}

For the $J_1$-$J_2$ Heisenberg model, the type of transition between QSL and VBS has long been a challenging problem. Our analysis of the degeneracy of the excitation near the phase transition shows that there is a four-fold degeneracy at the transition point, which arises from the crossing of a singlet state and a triplet state. Further analysis of the spectral weight reveals that the singlet excitation comes from dimer excitation at $(0,\pi)$, while triplet excitation comes from spin excitation at $(\pi,\pi)$. This momentum information aligns with previous VMC calculation~\cite{PhysRevX.11.031034}. The presence of a quadruplet state can be regarded as an implication of O(4) symmetry at the boundary of the QSL-VBS transition, which is identified by the recovery of the equivalence of spin-spin and dimer-dimer correlation in previous PEPS study~\cite{PhysRevX.12.031039, liu2023emergent}.  Most interestingly, we find there are two branches of dimmer singlet excitation crossing the triplet excitation near the transition point as shown in~\cref{fig: spectral_weight_W5_chi1024}. The four-fold degeneracy can be lifted to the five-fold degeneracy implying the possible emergent SO(5) symmetry discussed in~\cite{PhysRevX.12.031039, liu2023emergent} with increasing bond dimension $\chi$. The evidences are shown for small system widths ($W$) and sufficiently large bond dimension ($\chi$) in Appendix~\cref{fig: gap-J2_all}. However, we cannot ascertain if this five-fold degeneracy is an intrinsic feature of the current model. One reason is the current $2 \times 2$ unit cell ansatz can only capture the localized excitation. Future investigations may consider employing a larger unit cell, such as a $3\times3$ or $4\times4$ lattice, which may be more suitable for the system. Implementing a larger unit cell approach could potentially involve modifications to the unit cell algorithm, such as incorporating multi-column helixes, to manage computational costs effectively.

Another reason for the uncertain five-fold degeneracy is the 1D nature of the present ansatz. The approach exhibits 1D characteristics when $W$ is small. The finite-W effect will break the rotational symmetry. While the strategy outlined in~\cref{sec: Mitigate the finite size effect} can help mitigate this issue, there is still room for further improvement. Future research to elucidate the emerging symmetry could explore an anisotropic model as suggested in~\cite{liu2023emergent}, where the O(4) symmetry is identified along the boundary between the QSL and VBS phases.

\section{Acknowledgments}
We thank Yue-Shui Zhang, Qi Yang, Hao Xie, Qi Zhang, Shuang Liang, Hai-Jun Liao, Zheng-Cheng Gu, Shou-Shu Gong, Mingpu Qin, Anders Sandvik, Frank Verstraete, Jutho Haegeman and Laurens Vanderstraeten for the discussion. This project is supported by the National Key Projects for Research and Development of China Grant. No. 2021YFA1400400, National Natural Science Foundation of China (Grant Nos T2225018, 92270107, T2121001, 12188101), and Strategic Priority Research Program of the Chinese Academy of Sciences under Grant No. XDB30000000.

\appendix

\section{Mitigate the finite size effect}
\label{sec: Mitigate the finite size effect}

To mitigate the finite size effect, we approximate the original graph summation term~\cref{equ: exact_summation}:
\begin{equation}
    \begin{aligned}
    S^{B}_{\ \bar{B}}=&\left\{\sum_{j=1}^{\lceil W/2\rceil-1}\left(e^{ik_y \cdot j}\wang^{j-1}\right)+\frac{1+e^{ik_x}}{2}e^{ik_y \cdot \lceil W/2\rceil}\wang^{\lceil W/2\rceil-1} \right. \\
    &\left. +\sum_{j=\lceil W/2\rceil+1}^{W}\left(e^{ik_x}e^{ik_y \cdot j}\wang^{j-1}\right)\right\}\cdot \sum_{j=0}^\infty\left(e^{ik_x}\wang^W\right)^j
    \end{aligned}
    \label{equ: approximation_summation}
\end{equation}

\begin{figure}[H]
    \centering 
    \includegraphics[width=0.5\textwidth]{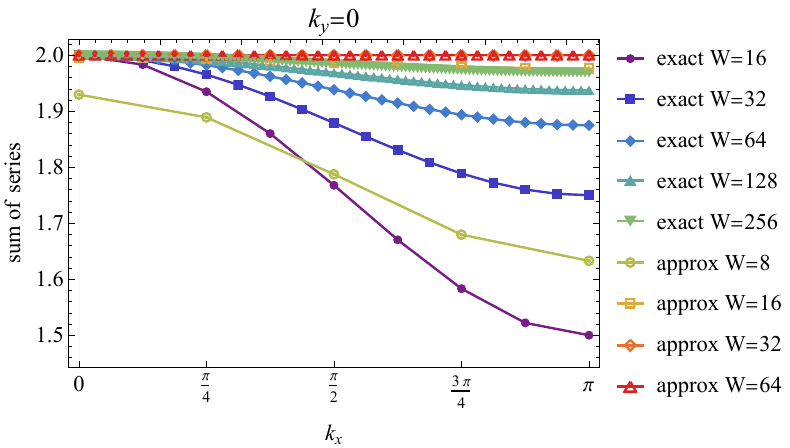}
    \caption{The comparison of the rate of convergence between exact and approximation for the first summation terms of $S^{B}_{\ \bar{B}}$ in \cref{equ: exact_summation} and \cref{equ: approximation_summation}. We take $k_y=0$ and  $\wang=0.5$ as a constant for example. For given $k_y$, the summation is not relative to $k_x$ in the infinite width limit.}
    \label{fig: approximation_rate_of_convergence}
\end{figure}

$\lceil \rceil$ is ceil int. $W-j$ and $je^{ik_x}$ represent the summation of this row and the next row respectively, and they compete with each other. We adopt a strategy where we use $W-j$ for the first $\lceil W/2\rceil-1$ terms, $je^{ik_x}$ for the last $\lceil W/2\rceil-1$ terms, and combine them in the middle term. Regardless of whether this approximation is employed, the first summation term for the $y$ direction of $S^{B}_{\ \bar{B}}$ converges to $\frac{e^{i k_y}}{1 - e^{i k_y} \wang}$ when $W \rightarrow \infty$. But the rate of convergence of approximation is faster, as shown in~\cref{fig: approximation_rate_of_convergence}.

\begin{figure}[H] 
    \centering
    \subfigure{
        \includegraphics[width=0.225\textwidth]{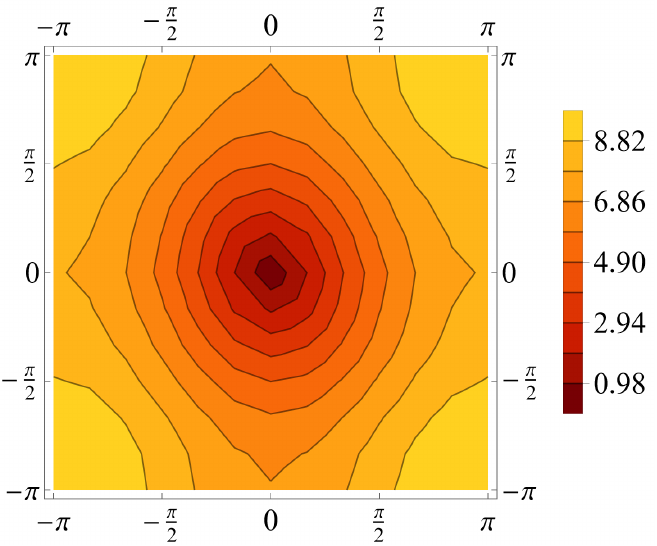}
        } 
    \subfigure{
        \includegraphics[width=0.21\textwidth]{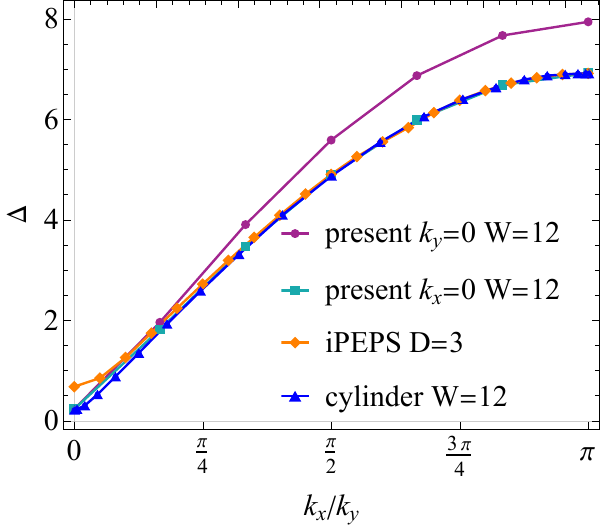}
        } 
    \subfigure{
        \includegraphics[width=0.225\textwidth]{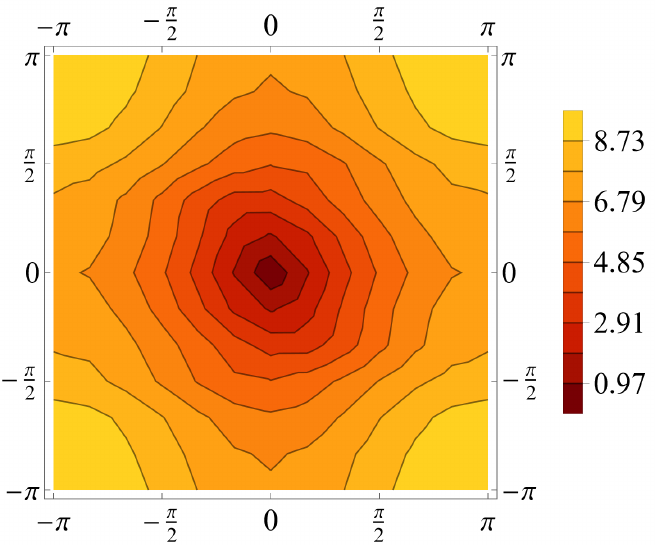}
        }
    \subfigure{
        \includegraphics[width=0.21\textwidth]{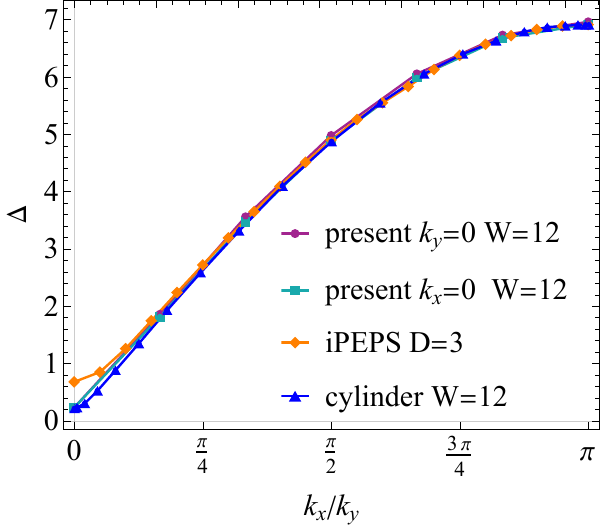}
        } 
    \caption{The top left panel (exact summation) and bottom left panel (approximate summation) display the first excitation gap $\Delta$ spectrum of the transverse field Ising model at the Brillouin zone ($W=12, \chi=256$). When performing exact summation based on~\cref{equ: exact_summation}, the spectrum exhibits C2 symmetry. In contrast, when employing approximate summation as outlined in~\cref{equ: approximation_summation}, the C4 symmetry is nearly recovered. The top right and bottom right panels provide a comparison between the results obtained from the cylinder geometry~\cite{PhysRevB.104.115142} and the iPEPS method. The comparison is made along the trajectory from $(0,0)$ to $(\pi,0)$ and $(0,0)$ to $(0,\pi)$ within the Brillouin zone.} 
    \label{fig: TFIsing_gap_brillouin_zone}
\end{figure}

In the helix geometry, only C2 symmetry is observed in the exact summation of the excitation spectrum. However, when using approximate summation, the C4 symmetry is almost recovered due to faster convergence. A detailed comparison with the results of the transverse field Ising model obtained on the cylinder geometry~\cite{PhysRevB.104.115142} demonstrates that this approximation successfully recovers the symmetry of the $x$ and $y$ directions. Hence, we employ this strategy for subsequent simulations and notice the excat and approximation are the same when $k_x=0$. 

\section{Magnetic order parameter of $J_1$-$J_2$ model}
\label{sec: magnetic order parameter of $J_1$-$J_2$ model}
\begin{figure}[H] 
    \centering
    \includegraphics[width=0.5\textwidth]{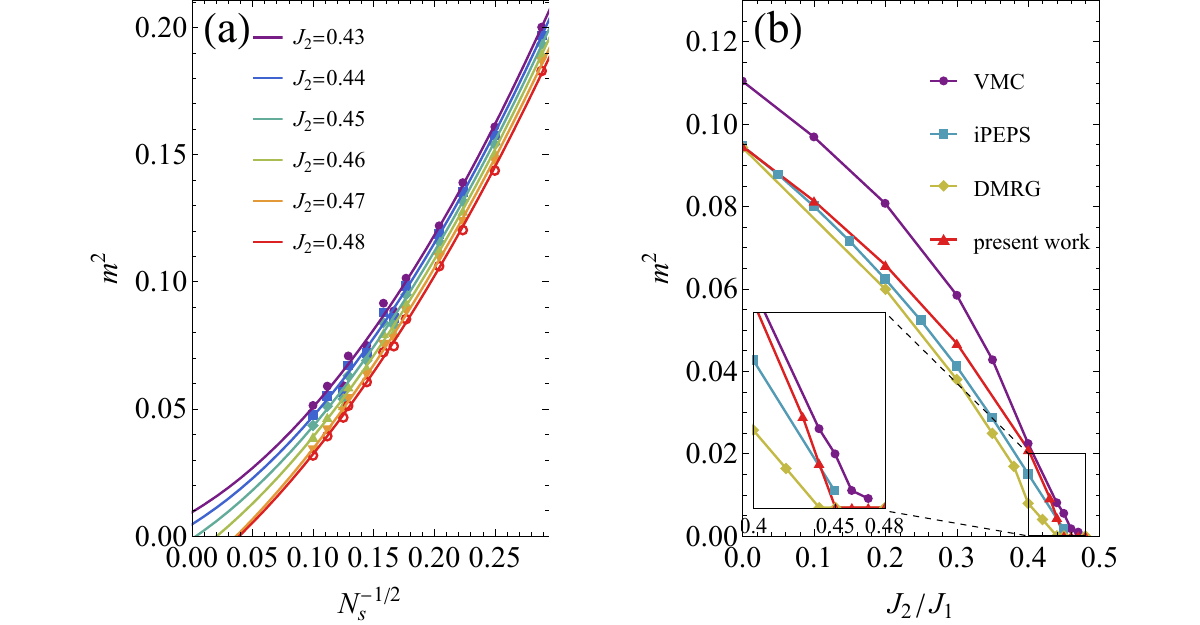}
    \caption{(a) $m^2$ plotted versus $N_s^{-1/2}$ (subregion size)  for infinite helix with $W=6,8,10$. (b) $m^2$ versus $J_2$ as obtained from (a).  For comparison, the data from infinite PEPS~\cite{10.21468/SciPostPhys.10.1.012}, finite cylinder DMRG~\cite{PhysRevLett.113.027201}, and VMC~\cite{PhysRevB.98.100405} are also included. The inset shows the region $0.4 < J_2 < 0.48$} 
    \label{fig: m2-J2}
\end{figure}

The Néel order parameter $m^2$ is defined as $m^2=\frac{1}{N^2} \sum_{i, j}\left\langle S_i \cdot S_j\right\rangle e^{i \boldsymbol{q} \cdot\left(\boldsymbol{r}_i-\boldsymbol{r}_j\right)}$, where $N$ is the total site number and $\boldsymbol{q}=(\pi,\pi)$. For infinite helix, we can't calculate $N\rightarrow \infty$ directly, so we calculate $m^2$ from the spin correlations of the subregion of $N_s=W \times 2$ to $W \times W$ sites on the infinite helix, which efficiently reduces boundary effects and enhances the dataset quality. In~\cref{fig: m2-J2}(a), we show $m^2$ for different systems with $W=6,8,10$ (bond dimension $\chi=1024$) and the subregion sizes $N_s$. We show the obtained two-dimensional limit $m^2_\infty$ by extrapolating the first and second term of $m_{N_s}^2=m^2_\infty+\alpha N_s^{-1 / 2}+\beta N_s^{-1}+\cdots$~\cite{PhysRevB.37.2380}. An overall plot $J_2$ dependence of the obtained magnetic order in the 2D limit $m^2_{\infty}$ compared with other methods is shown in~\cref{fig: m2-J2}(b). Such an analysis suggests that the Néel order vanishes for $J_2 > 0.45$, which aligns with previous MPS and PEPS calculations~\cite{PhysRevLett.113.027201, 10.21468/SciPostPhys.10.1.012, LIU20221034}. 

\section{Different size level crossing of $J_1$-$J_2$ model}

In the present analysis, we have examined various system widths, including $W=6, 8, 10, 12, 14$, and different bond dimensions denoted as $\chi=128, 256, 512$. We have consistently observed a level crossing point occurring around $J_2=0.52$. 

\begin{figure*} 
    \centering
    \includegraphics[width=1\textwidth]{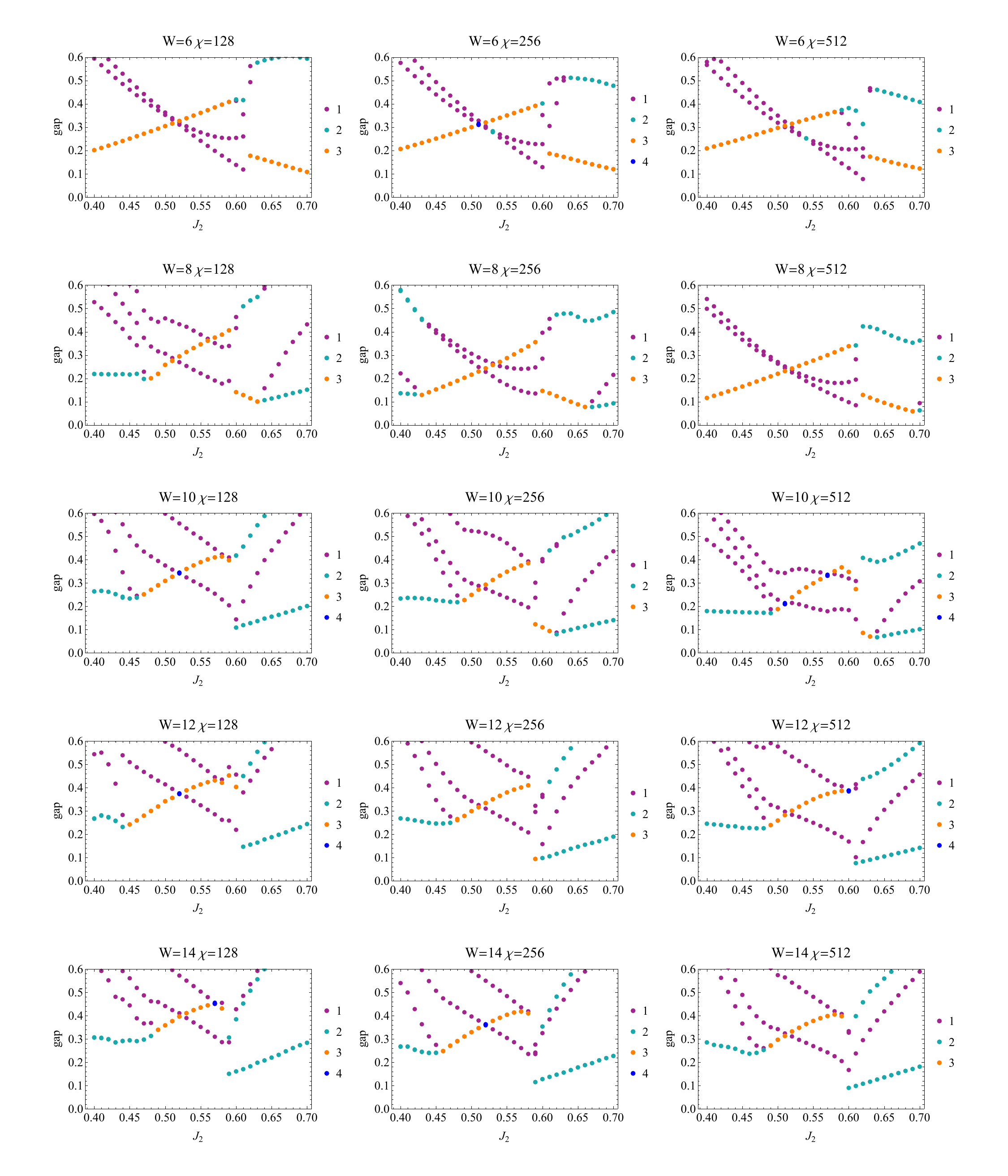}
    \caption{The energy gap along the $J_2$ axis is displayed in the figure, with various colors used to represent the degeneracy of the excitations. Notably, the transition point between QSL and VBS phases and VBS and stripe phases, occurring at approximately $J_2 \approx 0.52$ and $J_2\approx 0.61$, remains consistent across different system widths ($W$) and bond dimensions ($\chi$). The triplet excitations between the N\'eel and VBS phases appear with all simulations. But triplet excitations between the VBS and stripe phases disappear with large $W$ and small $\chi$, suggesting this is the finite-width effect of the 1D nature of the present ansatz.} 
    \label{fig: gap-J2_all}
\end{figure*}

We did not apply an extrapolation with $W$ as refs~\cite{PhysRevLett.121.107202, PhysRevX.11.031034, qian2023absence} because the bond dimensions $\chi$ is not large enough for different $W$ to converge. However, the level crossing is not sensitive to $\chi$ as shown in~\cref{fig: gap-J2_all}.

\newpage
\bibliographystyle{apsrev4-2} 
\bibliography{latex.bib}

\end{document}